\documentclass[onecolumn,english,epj]{svjour}
\usepackage[T1]{fontenc}
\usepackage[latin9]{inputenc}
\usepackage{amsmath}
\usepackage{amssymb}
\usepackage{graphicx}
\usepackage{setspace}
\usepackage{babel}
\begin{document}

\title{Comparing network covers using mutual information}

\author{Alcides Viamontes Esquivel\inst{1} \and  Martin Rosvall\inst{1}}

\institute{\inst{1}ICELab, Umeå University, Sweden}

\authorrunning{A. Viamontes Esquivel \and M. Rosvall}

\titlerunning{Comparing covers}

\abstract{In network science, researchers often use mutual information to understand
the difference between network partitions produced by community detection
methods. Here we extend the use of mutual information to covers, that
is, the cases where a node can belong to more than one module. In
our proposed solution, the underlying stochastic process used to compare
partitions is extended to deal with covers, and the random variables
of the new process are simply fed into the usual definition of mutual
information. With partitions, our extended process behaves exactly
as the conventional approach for partitions, and thus, the mutual
information values obtained are the same. We also describe how to
perform sampling and do error estimation for our extended process,
as both are necessary steps for a practical application of this measure.
The stochastic process that we define here is not only applicable
to networks, but can also be used to compare more general set-to-set
binary relations. }

\maketitle

\section{\label{sec:Introduction}Introduction}

Many complex phenomena can be characterized by interconnected networks
of basic parts: the network nodes. In order to gain in understanding
of these complex phenomena, researchers often start by grouping nodes
in non-overlapping modules, forming a so-called network \emph{partition}.
As there are many automatic ways of creating partitions from a raw
network, partitions of the same network generated by different methods
are frequently compared using one popular measure of similarity: the
mutual information (MI)\cite{lancichinetti2009benchmarks,barron1998minimum,fortunato,leung2009towards,barber2009detecting,ronhovde2009multiresolution,berry2011tolerating}.
But partitions are not always the most appropiate way of defining
functional components, as many phenomena can be better understood
if, instead of partitions, the more general structure of \emph{cover}
is allowed, where modules of the network can overlap or form nested
hierarchies. Researchers have developed methods that can automatically
detect those structures \cite{stabeler2011using,lancichinetti2011finding,schaub2011coding,kim2011map},
but carrying along the mutual information as a similarity measure
has proven more difficult, as we explain next.

When comparing two partitions, the mutual information is calculated
by taking all the pairs of modules $(x,y)$, one from each partition,
and \emph{counting} the number of common nodes that these modules
have. The count, divided by the total number of nodes in the network
and denoted as $p(x,y)$, is used together with the fraction of nodes
$p(x)$ and $p(y)$ that each of the modules $x$ and $y$ holds in
its respective partition. These fractions are disguised as probabilities
in the mutual information formula\cite{cover2006elements}:

\begin{equation}
I(X;Y)=\sum_{y\in Y}\sum_{x\in X}p(x,y)\log_{2}\left(\frac{p(x,y)}{p(x)p(y)}\right).\label{eq:discrete-mi-1-1}
\end{equation}
We identify this way of calculating the MI as the \emph{counting approach. }

\begin{figure*}
\centering\includegraphics[width=0.7\textwidth]{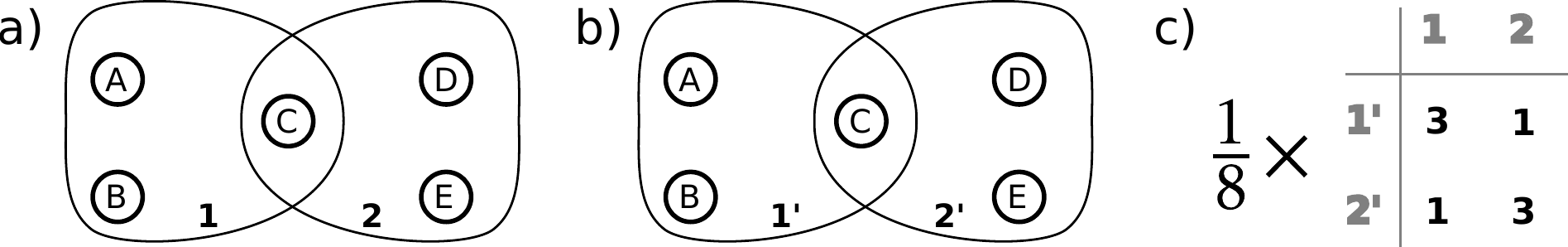}

\caption{\label{fig:Why-the-counting}Why the counting approach can not be
readily extended to covers. The cover in (a) is compared with itself
(b); the only difference between the two subfigures is in module names
($\mathbf{1}$ and $\mathbf{2}$ vs $\mathbf{1'}$ and $\mathbf{2'}$).
The number of common nodes for each pair of modules is shown in (c).
Despite these two covers being equivalent with regard to node memberships,
the mutual information (Eq.\,\ref{eq:discrete-mi-1-1}) calculated
with the fractions from the counting approach is not equal to Shannon's
entropy.}
\end{figure*}

We call two partitions or covers equivalent if they are different
only in the choice of module's names (Fig.\,\ref{fig:Why-the-counting}).
In the counting approach, if the two compared partitions are equivalent,
the mutual information reaches a maximum value equal to the Shannon's
entropy of the fractions $p(x)$ and $p(y)$: $I(X;Y)=H(X)=H(Y)$.
This happens because, for any two pair of modules $x$ and $y,$ either
they contain exactly the same nodes, and thus $p(x,y)=p(x)=p(y)\neq0$,
or they have no common nodes at all and $p(x,y)=0$. 

With covers, the problem is that it is not evident what to count:
the fractions used by the counting approach don't play well as probabilities
any longer. For example, if one insists on using the fractions, the
equality between MI and entropy won't be valid for two equivalent
covers (Fig \ref{fig:Why-the-counting}). 

Nonetheless, there are ways of comparing covers inspired by the mutual
information. For example, the authors of \cite{lancichinetti2009benchmarks}
see a node belonging to a module as a yes-or-no fact, independent
of everything else in the network and reflected on a binary vector.
Starting from the concept of mutual information, they use this vector
to ultimately produce a clustering-similarity measure. Being an approximation,
expressions equivalent to Eq.\,\ref{eq:discrete-mi-1-1} are used
indirectly. Consequently, when their procedure is applied to singly-assigned
nodes in partitions, the values obtained are not the same as in the
conventional counting approach.

Here we take another course. For us, if a node is in many modules,
we consider only one of those modules at a time, depending on an assumed
context, such that different contexts can yield different node-module
memberships. If we were talking about a person, for example, we would
implicitly consider her part of a particular social group, depending
on together with whom we mention her: a friend, a colleague, or a
relative. We use this context principle to derive a stochastic process
that disambiguates multiple memberships and yields module pairs, one
single module from each cover and context. The probabilities of these
pairs can be used straightforwardly in the mutual information formula
(Eq.\,\ref{eq:discrete-mi-1-1}). Thus, we don't require any changes
to the mutual information definition, and our results are compatible
with the counting approach. We call the obtained stochastic process
\emph{extended.}

The rest of this work is organized as follows. In section \ref{sec:The-mutual-information},
we re-introduce the counting approach as a simple stochastic process,
which justifies our intention of keeping it as a limit case when comparing
covers which happen to be partitions. In section \ref{sec:The-stochastic-process},
we extend this stochastic process to covers. We complement section
\ref{sec:The-stochastic-process} with an Appendix that introduces
the extended process using basic set theory, independet of the network
concepts that we use in the main part of the paper. Section \ref{sec:Behavior-of-the}
shows that our extended process is sensitive to several kinds of differences
between covers. Finally, in section \ref{sec:Simulation-and-error}
we explain how to control the error in the simulation procedure used
to calculate our new measure.

\section{\label{sec:The-mutual-information}Partitions and normalized MI}

Here we show how comparing partitions is reduced to comparing random
variables, and justify the counting approach conventionally used.
In this and the following section, we will be assisted by a metaphor.
In the metaphor, a \emph{caddie} is choosing nodes with uniform probability
from the network. Meanwhile, two \emph{players} are in possession
of each of the partitions. The players observe each node handled by
the caddie, and announce the module that the node belongs to, according
to their respective partition. The modules that each player reports
are taken as the random variables $X$ and $Y$, whose probabilities
we will use in the mutual information formula (Eq.\,\ref{eq:discrete-mi-1-1}).
Because the caddie is drawing nodes with uniform probability, the
probabilities of the random variables $X$ and $Y$ and the joint
probability of any given pair can be calculated exactly as the proportion
of nodes in each module and the fraction of nodes common to the pair
of modules, respectively. Therefore, the counting approach can be
framed as the MI of a stochastic process. We will call this stochastic
process \emph{conventional}, and we will keep it as a limit case occurring
when we compare partitions using our extended process. We will define
the extended process in the next section, but first, we will discuss
the issue of normalization. 

\global\long\def\hh{I_{a}}

\global\long\def\hm{I_{n}}

Normalization is required for getting a convenience indicator: one
where 0 means no similarity and the value 1 is special because it
means that the random variables are interchangeable, or, in our case,
that the partitions or covers are equivalent. There are two ways in
which normalization is normally done\cite{mcdaid2011normalized}.
In the first one, the mutual information is divided by the average
of the Shannon's entropies, while in the second one, it is divided
by the maximum of the entropies. We advocate for using the maximum
as the divisor because of the following. Suppose that four random
variables $X$,$Y$,$Z$ and $W$ have Shannon's entropies $H(X)=10$,
$H(Y)=15$, $H(Z)=10$ and $H(W)=2$ . Furthermore, suppose the MI
between $X$ and any other of the two variables is $2$. If the average
is chosen as the divisor, the average-normalized MI between $X$ and
the rest of the variables would be $\bar{I}_{a}(X;Y)=4/25$,$\hh(X;Z)=1/5$,
$\hh(X;W)=1/3$. The variable $W$ gets a higher score than all of
the others just because its entropy is lower, while $Z$ gets a higher
score than $Y$ just because its entropy is similar to $X$'s. In
other words, $Y$ is correctly penalized for overdescribing, but $W$
is wrongly rewarded for oversimplifying. Using the maximum entropy
as a divisor, on the other hand, we would get instead $\hm(X;Y)=2/15$,
$\hm(X;Z)=1/5$, and $\hm(X;W)=1/5$, which is more reasonable. Therefore
we take the maximum as the divisor:

\begin{equation}
I_{n}(X;Y)=\frac{I(X;Y)}{\max\left\{ H(X),H(Y)\right\} }\label{eq:max-normalized}
\end{equation}

\section{\label{sec:The-stochastic-process}The stochastic process for covers}

Here we look at how to extend the stochastic process defined in the
previous section to cases where a node can belong to multiple modules.
Appendix \ref{sec:The-mathematical-formulation} complements this
section with a shorter and conventional exposition of the same subject,
suitable for readers familiar with very basic set theory. 

\begin{figure*}
\centering\includegraphics[width=0.5\textwidth]{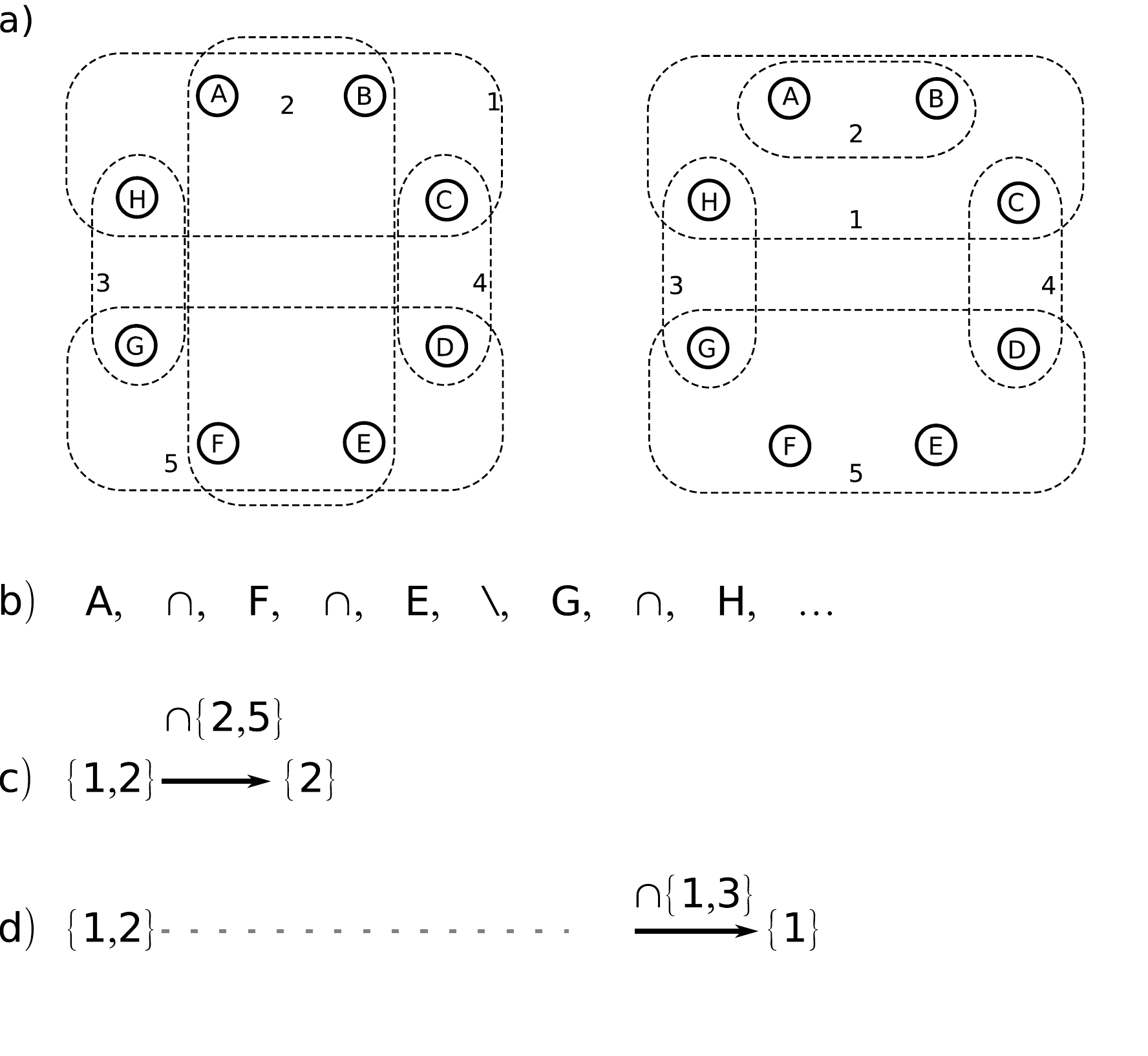}

\caption{\label{fig:Two-labelings_}Illustration of how the new stochastic
process works by looking at two example covers. Subfigure (a) shows
the two covers, made by a common set of nodes A,B,C,D,E,F,G,H. Subfigure
(b) shows an arbitrary interleaving of those nodes, alternating nodes
and set operations. Subfigures (c) and (d) show the set operations
outcome; the dotted portion of the line in subfigure (d) represents
steps that do not remove modules, while the operations that contribute
to the result are shown with a solid line. }
\end{figure*}

In covers, a node can be associated with more than one module. When
that happens, we use the context set by the memberships of other nodes
to highlight just one module of the first node at a time. This is
similar to a person associated with the groups \emph{family} and \emph{work}
singling out the group \emph{work} in the presence of colleagues.

In the players' metaphor that we introduced before, this disambiguation
mechanism allows the players to arrive at single modules, and, if
the covers are equivalent, at matching answers. Because each player
should be blind to what the other one does, he needs to base his actions
on common information that he receives from the caddie.

We call our particular way of structuring this common information
\emph{interleavings}. Interleavings are the simplest representation
of context that works with the membership relation between nodes and
modules in a cover. Each interleaving consists of an ordered sequence
of nodes, where between two consecutive nodes an operation bit is
inserted. This bit represents the choice between one of two set operations:
set intersection or set difference. Each node appears exactly once
in the interleaving, so, when looking to the nodes alone, we see a
permutation of nodes. As for the operation bits, they are randomly
chosen with equal probability: taken alone they look like a random
binary vector drawn from a uniform distribution. Figure \ref{fig:Two-labelings_}(b)
shows the first few elements of an example of interleaving, where
the nodes have been represented with uppercase letters and the operation
bits have been represented with the operation's usual mathematical
symbols. 

Let's see how a stochastic process can use the context provided by
an interleaving to arrive at a unique module of the cover. Now, instead
of handing nodes, the caddie produces an entire interleaving each
time. When each player sees the first node of the interleaving, he
will build a set with all the modules of that node, according to his
cover. In the example of Fig.\,\ref{fig:Two-labelings_}, both players
get the set $\{1,2\}$. These sets will need to be disambiguated using
the rest of the interleaving, and thus the players will keep them.
If a player's set contains only one module, he will output that module
and finish. Otherwise, the player will use the next set operation
and node from the interleaving. He uses the operation over his sets
of modules and the incoming node's set of modules. If the player gets
a non-empty result, he will replace his set with this result set.
In Fig.\,\ref{fig:Two-labelings_}(c) we see that the first player
can end almost inmediatly, using the modules that node $F$ has according
to the first cover . However, as Fig.\,\ref{fig:Two-labelings_}(d)
shows, the other player has to keep going until he executes the intersection
with the modules of node $H$. As long as no two modules in the same
cover share exactly the same nodes, the process always ends in such
a way that both players select a unique module. A proof of this is
shown in Appendix \ref{sec:Conditions-under-which}. 

From this definition of the extended process, it is straightforward
to use Eq.\,\ref{eq:max-normalized} to arrive at a value of the
mutual information. The most efficient way of obtaining the probabilities
for Eq.\,\ref{eq:max-normalized} is generating random interleavings
with the computer and doing the disambiguation process described above.
This is an approximate procedure, therefore, in section \ref{sec:Simulation-and-error}
we give additional details about how to perform the sampling and bound
the error.

In summary, the differences between the conventional process and the
extended one that we have introduced here, is that the first samples
one node at a time, while the second exploits the context created
by the rest of the node-module assignments in the network. As we hinted
before, interleavings are just one of many possible structures. If
a particular community detection method outputs more information associated
with nodes or modules, that information could probably be put to good
use. For example, it could influence the sampling of nodes or operations.

\section{\label{sec:Behavior-of-the}Behavior of the MI for cover differences}

In the following sub-sections, we consider only a very basic aspect
of the normalized mutual information (NMI): namely, that it is sensitive
to differences in covers due to partition structure, hierarchies and/or
overlaps.

\begin{figure}
\centering\includegraphics[width=0.4\columnwidth]{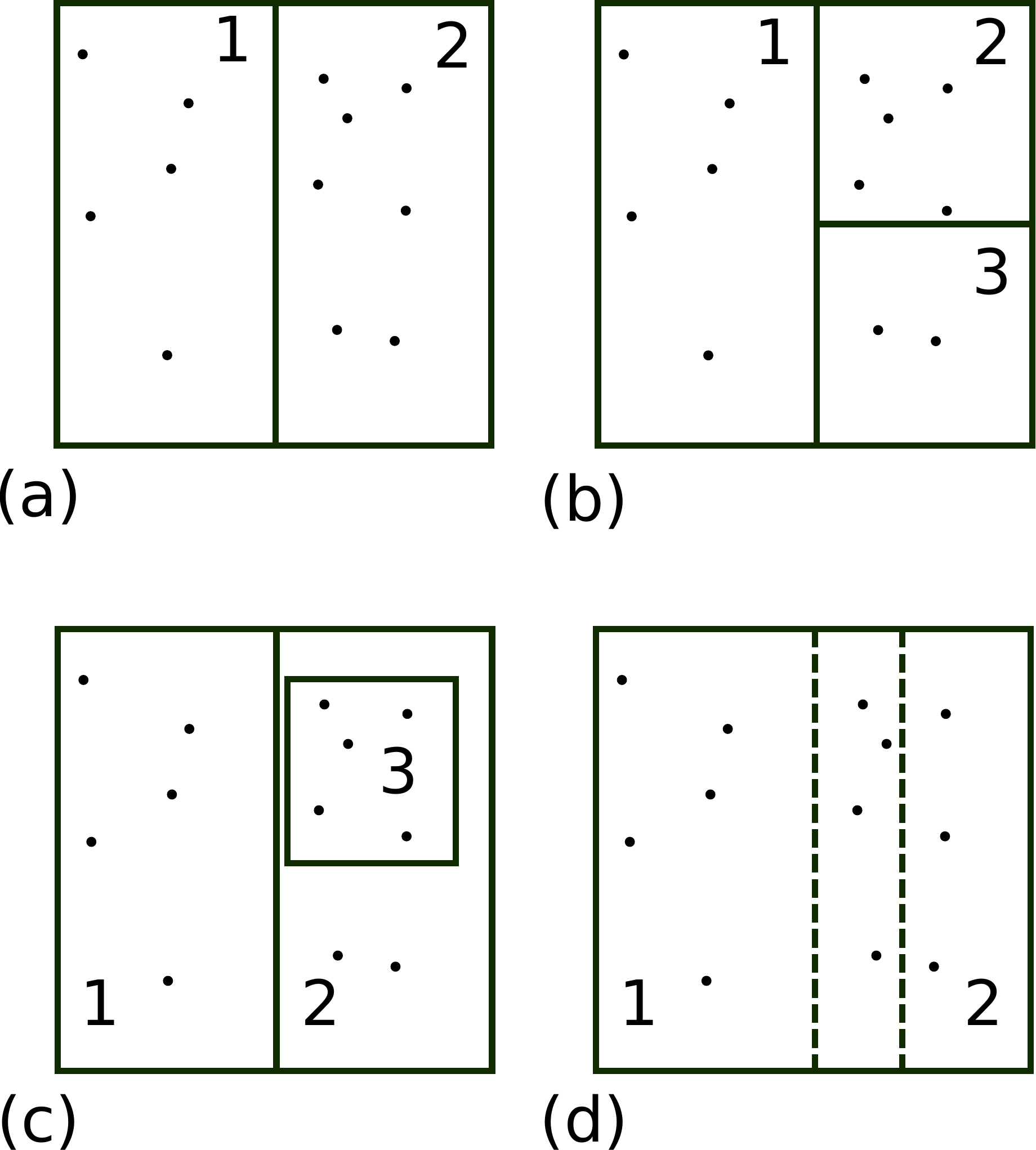}

\caption{\label{fig:Cases-an-matrices}Several example covers. The reference
cover in (a), is a partition made by two modules, denoted by 1 and
2, (b) shows a cover which is different just by splitting the previous
one's module 2 in two new modules: 2 and 3. In (c) a very simple hierarchy
is shown, with module 3 containing some of the nodes of module 2.
Finally, in (d) the nodes between the discontinuous line fringe belong
(overlap) to both module 1 and 2. The comparison of the first partition
with any of the other cases yields values of the NMI lower than 1. }
\end{figure}

The first thing to note is that, for any given partition $A$, its
normalized mutual information with itself, according to both the conventional
and the extended process will be 1. This property will remain valid
for covers in general using the extended process. 

Next we examine how this maximal value degrades when $A$ is compared
with a cover which is obtained from $A$ according to some process:
simple module division, introduction of a hierarchy, or introduction
of overlapping modules.

\subsection{Sensitivity to module splitting}

If two partitions are compared, the values of the mutual information
determined by both the conventional and the extended process are the
same. That's because the extended process will be able to decide an
output module using just the first node from each interleaving. Therefore,
the observations below will apply to both processes.

Given a partition $A$ , if $A'$ is obtained from $A$ by splitting%
\footnote{Splitting in a non-trivial way: the parts can not be empty.%
} one of its clusters (see Fig.\,\ref{fig:Cases-an-matrices} (a)
and (b) ), the mutual information between $A$ and $A'$ will be the
Shannon's entropy of $A$, because $A'$ determines $A$ completely;
thus $H(A|A')=0$ and $I(A;A')=H(A)-H(A|A')=H(A)$. However, normalization
according to Eq.\,\ref{eq:max-normalized} will give the value

\[
\frac{H(A)}{\max\{H(A),H(A')\}}
\]

and because $H(A')>H(A)$, the NMI will be less than 1.

\subsection{Sensitivity to introduction of hierarchies }

We consider the case where the cover $A'$ is obtained from $A$ by
creating a new module with a subset of nodes, all of which are also
already part of another module. In Fig.\,\ref{fig:Cases-an-matrices},
an example would be subfigures (a) and (c). Because this case considers
multiple modules per node, in this and the following sections we will
be only speaking of the extended process. 

We show that the comparison of $A$ and $A'$ will result in decreased
NMI. One of the conditions for an NMI of less than 1 is the one we
exploited in the previous section: that the random variables have
different entropy. 

Let's consider the outcome of the stochastic process for nodes of
module 3 (and thus $1$) in Fig.\,\ref{fig:Cases-an-matrices} (c).
These nodes belong to two modules, so the disambiguation process will
consume more of the interleaving, and will yield either module $3$
or module $1$. In all those cases, the cover in Fig.\,\ref{fig:Cases-an-matrices}(a)
will still yield module $1$. So, the cover with hierarchies will
produce a random variable with greater entropy than the partition
in Fig.\,\ref{fig:Cases-an-matrices}(a). However, because by looking
to Fig.\ref{fig:Cases-an-matrices}\,(c) one can always predict the
outcome of the process for (a), these two covers again have the same
mutual information, and the normalization in Eq.\,\ref{eq:max-normalized}
will yield a value minor than 1. Therefore, the structure of the random
variables for the extended process penalizes differences in hierarchies.

\subsection{Sensitivity to introduction of overlaps}

For overlaps, the mutual information itself is reduced: if we use
the extended process over the covers in Fig.\,\ref{fig:Cases-an-matrices}(a)
and Fig.\,\ref{fig:Cases-an-matrices}(d), upon seeing a node that
belongs to module 2 in Fig.\,\ref{fig:Cases-an-matrices}(d), it
is not possible to determine unequivocally what will be the resulting
module for the cover in Fig.\,\ref{fig:Cases-an-matrices}(a). Thus,
the MI between the two covers is less than the entropy of the partition
in Fig.\,\ref{fig:Cases-an-matrices}(a), and the obtained NMI value
is less than 1.

\section{Simulation and error control\label{sec:Simulation-and-error}}

\global\long\def\thefar{\theta_{i}^{\textrm{far}}}

The number of interleavings for the extended process grows very fast
with the number of nodes and modules, so it is no longer practical
to evaluate exactly the proportions of possible outcomes. But we can
actually do the simulation with a computer program in an efficient
way. That is, we can generate random interleavings and apply the disambiguation
procedure described in section \ref{sec:The-stochastic-process} and
Appendix \ref{sec:The-mathematical-formulation}, getting as many
pair of modules as needed to reach a good estimate of the NMI. When
we do the simulation process, the most likely outcomes will be, by
definition, the ones with bigger contributions to the mutual information
matrix $\left\{ p(x,y)\right\} _{x\in X,y\in Y}$, and this will help
us to bound the error. 

We will call each individual act of choosing a random interleaving,
applying the disambiguation procedure to it and getting a pair of
modules an \emph{event. }As explained before, we will count together
events that yield the same pair of modules.

The NMI, as calculated by Eq.\,\ref{eq:max-normalized}, is a function
of the joint probabilities of events, denoted by $p(x,y)$. Also,
the marginals $p(x)$ and $p(y)$, and the entropies $H(X)$ and $H(Y)$,
are calculable from the joint probabilities. Because the joint probabilities
are the basic input to all the calculations, it is convenient to rename
them introducing variables $\theta_{1},\ldots\theta_{m}$ . Each $\theta_{i}$
denotes the true probability of an event yielding a particular pair
of modules. This way, it is possible to write $I_{n}$ as a function
application result $I_{n}=f\left(\theta_{1},\ldots\theta_{m}\right)$,
where $f$ is given by equations \ref{eq:discrete-mi-1-1} and \ref{eq:max-normalized}. 

We don't have any of the true probability values $\theta_{i}$. Instead,
we count the number $\zeta_{i}$ of events corresponding to $\theta_{i}$
, and the total number of events $N$ during the simulation: $N=\sum_{i=1}^{m}\zeta_{i}$.
This count allows us to evaluate $f$ using frequencies, and obtain
an estimate $\tilde{I}_{n}=f\left(\frac{\zeta_{1}}{N},\ldots\frac{\zeta_{m}}{N}\right)$
of the NMI. Because the value $\tilde{I}_{n}$ is a random variable,
we can at best give a probabilistic estimate of the error $\left|\tilde{I}_{n}-I_{n}\right|$
. That is, given a risk $\epsilon$ and an error tolerance $e$, we
simulate as many events $N$ as needed for

\begin{equation}
\textrm{Pr}\left(\left|\tilde{I}_{n}-I_{n}\right|<e\right)\ge1-\epsilon\label{eq:big_i_prob-1}
\end{equation}

We need to relate the error $\left|\tilde{I}_{n}-I_{n}\right|$ with
$\zeta_{1},\ldots,\zeta_{m}$ and $N$. If we consider $f$ as approximately
linear inside of a small hypercube $\eta$ centered at $(\theta_{1},\ldots,\theta_{m})$
and spanning $\left|\theta_{i}-\frac{\zeta_{i}}{N}\right|$ in each
dimension, we can get an approximation for $\left|\tilde{I}_{n}-I_{n}\right|$.
Using first-order Tylor expansion and, in order to arrive at a worst
case, taking absolute values, we get:

\begin{equation}
\left|\tilde{I}_{n}-I_{n}\right|\leq\sum_{i=1}^{m}\left|\frac{\partial I_{n}}{\partial\theta_{i}}\cdot\left(\frac{\zeta_{i}}{N}-\theta_{i}\right)\right|\label{eq:uncertainty-1}
\end{equation}
In practice we replace$\left|\frac{\partial I_{n}}{\partial\theta_{i}}\cdot\left(\theta_{i}-\frac{\zeta_{i}}{N}\right)\right|$
by a finite difference equivalent: 
\begin{multline}
\frac{\partial I_{n}}{\partial\theta_{i}}\cdot\left(\frac{\zeta_{i}}{N}-\theta_{i}\right)\le\frac{\partial I_{n}}{\partial\theta_{i}}\cdot\left(\frac{\zeta_{i}}{N}-\thefar\right)\approx\\
f(\frac{\zeta_{1}}{N},\ldots,\frac{\zeta_{i}}{N},\ldots\frac{\zeta_{m}}{N})-f(\frac{\zeta_{1}}{N},\ldots,\thefar,\ldots\frac{\zeta_{m}}{N})\label{eq:fd-approx}
\end{multline}
where instead of $\theta_{i}$ we use a \emph{worst-case} probability
$\thefar$. We say that this is a worst-case probability because it
is the probability value farthest from $\frac{\zeta_{i}}{N}$ that
would make the count $\zeta_{i}$ probable enough to satisfy the risk
condition (Eq.\,\ref{eq:big_i_prob-1}).

To obtain $\thefar$, we exploit the fact that each count $\zeta_{i}$
is approximately distributed following a binomial around the true
probability: $\zeta_{i}\sim B\left(N,\theta_{i}\right)$. We assume
a so-called \emph{component} \emph{risk} $\xi$, which we will shortly
link to $\epsilon$, and solve the inverse problem of finding a $\thefar$
such that:

\begin{multline}
\textrm{Prob}\left(X\le\zeta_{i}\right)=\xi/2\\
\textrm{ given that }X\sim B(N,\theta_{i}^{\textrm{far}}).\label{eq:inverse-problem-1}
\end{multline}

This way, getting an event count in the interval between $\zeta_{i}$
and $N-\zeta_{i}$ will have probability $1-\xi$, according to $B(N,\theta_{i}^{\textrm{far}})$.
When we account for all the variables $\zeta_{1},\ldots\zeta_{m}$,
the probability of being inside the hypercube $\eta$ will be $(1-\xi)^{m}$.
Equating this with the complement of our risk 
\[
1-\epsilon=(1-\xi)^{m}
\]
and then doing first-order Taylor approximation assuming that $\epsilon$
is small enough, we obtain $\xi\approx\epsilon/m$. We can use this
value of $\xi$ to solve for $\thefar$ in Eq.\,\ref{eq:inverse-problem-1}.
Then, substituting upwards in Eq.\,\ref{eq:fd-approx} and then in
Eq.\,\ref{eq:uncertainty-1} we get the error's upper bound. In this
way, all that we need is to simulate enough events as for this error's
upper bound to fall below $e$.

\section{Conclusions}

We have defined a stochastic process that extends the use of mutual
information to compare covers of networks and that in the limit case
of network partitions yields the same results of the conventional
process for partitions. 

A notable characteristic of our extended process is that it only uses
the membership relationships between nodes and modules. As appendix
\ref{sec:The-mathematical-formulation} shows, the extended process
can be stripped from the network terminology and introduced instead
using mathematical set-algebra. In other words, our extended process
is not restricted to networks, but can be used to compare two binary
set relations with a common finite domain. 

There are many ways of defining similarity, and a single-number measure
like ours can not possibly encompass all of them. Depending on what
features researchers want to highlight when making comparisons, different
measures must come into play. Nevertheless, the methodology of constructing
random variables aware of the features that need to be compared, and
then comparing the random variables using mutual information, can
be adapted to more specific requirements. 

We have made the source code of our implementation available at \emph{http://www.tp.umu.se/\textasciitilde{}alcides/}.
\begin{acknowledgement}
We were supported by the Swedish Research Council grant 2009-5344.
\end{acknowledgement}
\appendix

\section{\label{sec:The-mathematical-formulation}Formal definition of the
extended process \label{sub:The-definition-of}}

This appendix extends the exposition in section \ref{sec:The-stochastic-process}
with formal notations and definitions.

The term \emph{{}``}cover\emph{'' }that we used in the main text
is just an alias for \emph{binary relation. }That is, a cover $M$
is a binary relation between a finite set of {}``nodes'' $E$ and
a set of {}``modules'' $L$. Using set notation: $M\subset E\times L$,
where {}``$\times$'' denotes Cartesian product. Given a node $e\in E$,
we will use the term $\ell_{M}(e)$ to refer to the set of modules
associated with $e$. That is, a module $l\in\ell_{M}(e)$ if and
only if $(e,l)\in M$.

We define an \emph{interleaving} $i_{E}$ of $E$ as the ordered sequence
\[
i_{E}=[e_{0},b_{1},e_{1},\ldots,b_{\left\Vert E\right\Vert -1},e_{\left\Vert E\right\Vert -1}],
\]
 where $[e_{0},e_{1},\ldots e_{\left\Vert E\right\Vert -1}]$ forms
a permutation of $E$ (that is, all the elements of $E$ in a particular
order) and $b_{i}$ is either $0$ or $1$. We reserve the capital
$I_{E}^{+}$ to denote the set of all posible interleavings with elements
of $E$. Note that the cardinality of this set is $\left(\left\Vert E\right\Vert \right)!\times2^{\left\Vert E\right\Vert -1}$. 

For any given $M\subset E\times L$, we define a function $J_{M}$
that takes an interleaving and returns a subset of modules: $J_{M}:I_{E}^{+}\rightarrow\mathcal{P}(L)$.
Here $\mathcal{P}(\cdot)$ denotes the power set of {}``$\cdot$''.
We define $J_{M}\left([e_{0},b_{1},e_{1},\ldots,b_{\left\Vert E\right\Vert -1},e_{\left\Vert E\right\Vert -1}]\right)$
as the result $S_{\left\Vert E\right\Vert }$ of applying the following
procedure: start with $S_{0}=\ell_{M}(e_{0})$. If $S_{k}$ has been
calculated, calculate $S_{k+1}$ in the following way, for $k=0,1,2\ldots,\left\Vert E\right\Vert -1$
\begin{equation}
S_{k+1}=\begin{cases}
s* & \textrm{if }s*\nsubseteq\emptyset\textrm{ and }b_{k}=1\\
s^{-} & \textrm{if }s^{-}\nsubseteq\emptyset\textrm{ and }b_{k}=0\\
S_{k} & \textrm{otherwise}
\end{cases}\label{eq:sk_evolution}
\end{equation}

where
\[
s*=S_{k}\bigcap\ell_{M}(e_{k})
\]
and
\[
s^{-}=S_{k}\setminus\ell_{M}(e_{k})
\]

Note that $J_{M}\left([e_{0},b_{1},e_{1},\ldots,b_{\left\Vert E\right\Vert -1},e_{\left\Vert E\right\Vert -1}]\right)\subseteq\ell_{M}(e_{0})$.
For our definition, we don't need the entire set $I_{E}^{+}$, but
rather the subset $I_{E,M}$ such that $i\in I_{E,M}$ if and only
if $\left\Vert J_{M}(i)\right\Vert =1$. If an element $i$ belongs
to $I_{E,M}$, we say that $J_{M}$ is \emph{well defined} for $i$.
In the next section we discuss under which conditions $I_{E}^{+}$
is equal to $I_{E,M}$.

Now it is straightforward to define the extended stochastic process:
given two covers $M_{1}\subset E\times L_{1}$ and $M_{2}\subset E\times L_{2}$
over the same set of elements $E$, we assume the existence of a random
variable with values over $I_{E}=I_{E,M_{1}}\bigcap I_{E,M_{2}}$
, such that all values of $I_{E}$ have the same probability. We can
apply the functions $J_{M_{1}}$ and $J_{M_{2}}$ over $I_{E}$. The
result are the modules $J_{M_{1}}(I_{E})$ and $J_{M_{2}}(I_{E})$,
which in turn can be considered random variables over the set of modules
$L_{1}$ and $L_{2}$. We define the MI of the two covers $M_{1}$
and $M_{2}$ as the mutual information between $J_{M_{1}}(I_{E})$
and $J_{M_{2}}(I_{E})$.

\section{\label{sec:Conditions-under-which}Conditions under which the extended
process yields only one module as response}

The function $J_{M}$ defined in Appendix \ref{sec:The-mathematical-formulation}
may yield either a set with one element, or a set with more elements.
In this Appendix, we prove that if every pair of modules of the cover
is different regarding the nodes they contain, then the function $J_{M}$
will yield a set with just one element.

First we observe that the sequence of intermediate results defined
by Eq.\,\ref{eq:sk_evolution} and corresponding to a particular
interleaving $[e_{0},b_{1}\ldots,b_{\left\Vert E\right\Vert -1},e_{\left\Vert E\right\Vert -1}]$
is non-increasing: $S_{k+1}$ will either have as many elements as
$S_{k}$ or fewer, and evidently, $S_{k+1}\subseteq S_{k}$. Now we
examine a final set $F=S_{\left\Vert E\right\Vert }$, whose cardinality
is greater than 1. Because the family of sets $S_{k}$ is non-increasing,
we have that $F\subset S_{k}$ for $k=0,1,\ldots,\left\Vert E\right\Vert -1$.
If the set $F$ was conserved in all the $\left\Vert E\right\Vert -1$
operations prescribed by Eq.\,\ref{eq:sk_evolution}, it means that:
\begin{description}
\item [{for~set~intersection~operations:}] $F$ was always present in
the set $\ell_{M}(e_{k})$ for $k=1,\ldots e_{\left\Vert E\right\Vert -1}$
and thus not eliminated, or it was not present at all (in any of its
members), and the operation was discarded because it resulted in the
empty set.
\item [{for~set~difference~operations:}] $F$ was never present in the
set $\ell_{M}(e_{k})$ for $k=1,\ldots e_{\left\Vert E\right\Vert -1}$
and thus not eliminated, or it was present but the operation was discarded
because it resulted in an empty set. That means that all the elements
of $F$ were in $\ell_{M}(e_{k})$. 
\end{description}
These two branches converge in the fact that the set $F$ was always
present or absent as a whole in each of the sets $\ell_{M}(e_{k})$
, and because each interleaving contains all the nodes, every module
present in the set $F$ had exactly the same nodes. 

\bibliographystyle{/usr/share/texmf/bibtex/bst/epj/epj}
\bibliography{hierarchies}

\end{document}